

\font\twelvebf=cmbx10 scaled \magstep1
\hsize 15.2true cm
\vsize 22.0true cm
\voffset 1.5cm
\nopagenumbers
\headline={\ifnum \pageno=1 \hfil \else\hss\tenrm\folio\hss\fi}
\pageno=1

\hfill IUHET-306

\hfill  June 26, 1995
\medskip
\bigskip
\centerline{\twelvebf Quantitative corrections to mass sum
rules}
\centerline{\twelvebf  involving baryons containing
heavy quarks}

\medskip
\bigskip

\centerline{D. B. Lichtenberg}

\centerline{Physics Department, Indiana University,
Bloomington, IN 47405, USA}

 \vskip 1 cm

Quantitative corrections are estimated to three of
Franklin's sum rules involving the masses of baryons
containing at least one charmed quark and
to three analogous sum rules for baryons containing
at least one bottom quark. The corrections
arise from three-body contributions to baryon interaction
energies and are calculated from a semiempirical
formula for the colormagnetic contributions to
baryon masses.


\bigskip
\medskip \bigskip \bigskip

About 20 years ago, Franklin [1] derived sum rules
relating the masses of baryons containing
at least one charmed quark. Most of these
baryon mass relations have not yet been tested
by experimental measurements.
Franklin restricted himself to ground-state baryons
without radial or orbital excitations, and so do we
in this work.

The Franklin sum rules are based on the
constituent quark model plus the assumption that
the interaction energy of quarks in a baryon consists of
only one- and  two-body terms. However, according to
quantum mechanics, the interaction
energy should contain a three-body term even if the
forces between quarks consist only of two-body
terms [2--5]. These three-body terms in baryons give
rise to corrections to Franklin's sum rules. It
is the purpose of this Brief Report
to estimate the magnitude of these corrections
in certain selected cases.

Franklin's sum rules fall into two categories: (i)
formulas involving the mass splittings of
baryon isospin multiplets, and
(ii) formulas relating the masses of different isospin
multiplets. In this work we neglect isospin breaking
and consider only mass relations of type (ii).
Furthermore, our method allows us to calculate
corrections to only three of Franklin's
formulas in category (ii). These three formulas
involve mass differences among baryons with spin 3/2
and spin 1/2 with the same quark content.
These mass differences arise from spin-dependent forces
and, to a good approximation,
are caused by the colormagnetic interaction of QCD [5].
We also obtain corrections to three analogous formulas
in which $c$ quarks are replaced by $b$ quarks.

For some baryons, the spin-dependent mass splittings
are known from experiment [5]; in other cases the
splittings can be estimated from a semiempirical mass
formula [6--8]
with parameters chosen so as to get a best
fit to the observed splittings.
In the case in which the splittings are known from
experiment, the magnitudes
of three-body interaction energies
have been deduced [5]. The semiempirical mass formula
incorporates some three-body effects
as a built-in feature. As a consequence, we can use the
semiempirical formula to get a good estimate of the
magnitude of the breaking of Franklin's formulas
relating the spin splittings in different baryons.

In the following, we let the symbol for a baryon
denote its mass. We generally follow
the notation of the Particle Data Group [9], with the
following amplifications. An asterisk on the symbol
for a baryon means that it has spin 3/2.
If a baryon contains
two quarks of the same flavor, they are the first two.
If all three quarks have different flavors, the two
lightest are the first two. In the latter case, there
are two inequivalent spin-1/2 states. One of these
states may be chosen so that the first two quarks have
spin 1 and the other so that the first two quarks have
spin 0. It has been shown [10] that in the physical
baryons the mixing is small
between the two ideal states of spin
1/2 defined above, provided the first two quarks
are the two lightest ones.
If the symbol [9] for the two spin-1/2 baryons
is otherwise the same, we put a
prime on the state in which the lightest two quarks have
spin 1. We let $q$ stand for either a $u$ or $d$ quark.

We neglect isospin breaking terms. Then
we can take linear
combinations of Franklin's sum rules [1] to obtain
the following relations among baryon masses:
$$\Sigma_c^*(qqc) - \Sigma_c(qqc) =
\Xi_{cc}^*(ccq) - \Xi_{cc}'(ccq), \eqno(1)$$
$$\Omega_c^*(ssc) - \Omega_c(ssc) =
\Omega_{cc}^*(ccs) - \Omega_{cc}(ccs), \eqno(2)$$
$$\Sigma_c^*(qqc) - \Sigma_c(qqc)
+\Omega_c^*(ssc) - \Omega_c(ssc) =
2[\Xi_c^*(qsc) - \Xi_c'(qsc)], \eqno(3)$$
where the symbol for a baryon is followed by its
quark content in parentheses. In subsequent
equations, we omit repeating
the quark content of these baryons.

Franklin did not consider sum rules for baryons
containing $b$ quarks, as the $b$ quark had not yet
been discovered when he wrote his paper.
However, as is clear from his paper,
equations analogous to (1)--(3) hold for $b$-quark baryons
simply by the replacement of all $c$ quarks by $b$ quarks
in the formulas.

In  Ref.\ [5], the following expressions were derived
relating the masses of baryons with different spin
wave functions but having the same quark content:
$$B^*(123)- B'(123) = 3R(132) + 3R(231), \eqno(4)$$
$$2B^*(123)+ B'(123)-3 B(123) = 12 R(123) \eqno(5)$$
where the $R(ijk)$ are three-body colormagnetic
interaction energies. If the first two quarks have
the same flavor, the prime should be omitted on
$B'(123)$ in Eq.\ (4), and Eq.\ (5) is absent.
Our notation differs
from that in Ref.\ [5] in several respects, including
the fact that we use the prime on the baryon whose
first two quarks have spin 1, whereas in [5] the
prime is used for the baryon whose first two quarks
have spin 0. If we have the values of $R(132)+R(231)$
for different baryons, we can use these values
to obtain corrections to Eqs.\ (1)--(3).
In order to correct the sum rules (1)--(3), we
need Eq.\ (4) but not Eq.\ (5).

For some baryons, the values of $R(132)+R(231)$ can
be obtained from experiment [5]. For other baryons,
the experimental data are not available. In the latter
cases, we use a semiempirical mass formula for
the spin splittings in baryons
in the form given in Ref.\ [7]. Application
of this formula leads to the values of $R(ijk)$ given in
Table I. For completeness, we have included the values of
$R(123)$ in Table I, even though these values are not
necessary for our considerations. The values of the
$R(ijk)$ in Table I are rounded to the nearest MeV.
We have also included in Table I the values of
$R(ijk)$ from Ref.\ [5] where the data are known.

The interaction energies $R(ijk)$ result from the
colormagnetic force between quarks $i$ and $j$.
In deriving his sum rules, Franklin assumed that the
$R(ijk)$ depend only on the flavors of the
quarks $i$ and $j$, but not on the flavor of the
``spectator'' quark $k$. However, it can be seen from
the values in Table I that there is a small dependence
on the flavor of the spectator quark. It is this dependence
that leads to corrections to Franklin's sum rules.
Using the values of $R(ijk)$ from the
semiempirical mass formula rather than the rounded
values in Table I, we obtain
the following formulas instead of Eqs.\ (1)--(3):
$$\Xi_{cc}^* - \Xi_{cc}'-
\Sigma_c^* + \Sigma_c = 15\ {\rm MeV}, \eqno(6)$$
$$\Omega_{cc}^* - \Omega_{cc}-
\Omega_c^* + \Omega_c = 10\ {\rm MeV}, \eqno(7)$$
$$\Sigma_c^* - \Sigma_c +\Omega_c^* - \Omega_c -
2(\Xi_c^* - \Xi_c')= 0, \eqno(8)$$
where we have rounded to the nearest 5 MeV.
These formulas give the corrections to Franklin's
formulas (1)--(3).
We see from the above equations that three-body
interaction energies lead to small corrections
in two of the sum rules and the third
survives essentially unchanged.

The corrections to Franklin's sum rules are still
smaller in the case of baryons containing $b$ quarks.
Although it may be a long time before mass
formulas for baryons containing $b$ quarks
can be tested, we nevertheless give these sum rules below:
$$\Xi_{bb}^* - \Xi_{bb}'-
\Sigma_b^* + \Sigma_b = 10\ {\rm MeV}, \eqno(9)$$
$$\Omega_{bb}^* - \Omega_{bb}-
\Omega_b^* + \Omega_b = 5\ {\rm MeV}, \eqno(10)$$
$$\Sigma_b^* - \Sigma_b +\Omega_b^* - \Omega_b -
2(\Xi_b^* - \Xi_b') = 0, \eqno(11)$$
again rounding to 5 MeV.

Although the semiempirical mass formula reproduces
the observed
spin-dependent splittings rather well in baryons
containing no heavy quarks, it is not known how
accurately it gives the splittings in baryons
containing heavy quarks, as the necessary
measurements have not yet
been carried out. However, even if the
formula is in error by 10  MeV or more,
Eqs.\ (6)--(11) should have much smaller errors;
we estimate these errors to be less than 5 MeV.
The reason is that these equations involve only
differences in spin-dependent splittings, and
therefore systematic errors in the
semiempirical mass formula will tend to cancel.

In conclusion, taking into account three-body interaction
energies leads to  small, but significant departures
from Franklin's baryon sum rules. In principle, these
effects can be tested by future measurements of
baryon masses.

\medskip

It is a pleasure to thank Jerry Franklin for valuable
discussions. This work was supported in part by
the Department of Energy.

\bigskip

\noindent References
\bigskip

\item{[1]} J. Franklin, Phys.\ Rev.\ D {\bf 12}, 2077 (1975).

\item{[2]} I. Cohen and H. J. Lipkin, Phys.\ Lett.\ {\bf
106B}, 119 (1981).

\item{[3]} J. M. Richard and P. Taxil,
Ann.\ Phys.\ (N.Y.) {\bf 150}, 267 (1983).

\item{[4]} D. B. Lichtenberg, Phys. Rev.
D {\bf 35}, 2183 (1987); Phys.\
Rev.\ D {\bf 40}, 3675 (1989).

\item{[5]} M. Anselmingo, D. B. Lichtenberg, and E.
Predazzi, Z.\ Phys.\ C {\bf 48}, 605 (1990).

\item{[6]} Yong Wang and D. B. Lichtenberg, Phys. Rev.
D {\bf 42}, 2404 (1990).

\item{7]} R. Roncaglia, A. Dzierba, D. B. Lichtenberg,
and E Predazzi, Phys.\ Rev.\ D {\bf 51}, 1248 (1995)

\item{[8]} R Roncaglia,  D. B. Lichtenberg, and E.
Predazzi, Phys.\ Rev.\ D {\bf 52}, Aug (1995)

\item{[9]} Particle Data Group: L. Montanet et al.,
Phys.\ Rev.\ D {\bf 50}, 1171 (1994).

\item{[10]} J. Franklin, D. B.  Lichtenberg,
W. Namgung, and D. Carydas,
Phys.\ Rev.\ D {\bf 24}, 2910 (1981).

\vfill\eject

Table I. Contributions to
baryon masses arising from the colormagnetic interaction.
The values of the colormagnetic interaction energy
$R(ijk)$ are
obtained from a baryon semiempirical mass formula
given in Ref.\ [7]. For comparison, where known, we give in
parentheses the values of $R(ijk)$ using experimental data
as input with the prescription of Ref.\ [5].

\vskip 12pt
$$\vbox {\halign {\hfil #\hfil &&\quad \hfil #\hfil \cr
\cr \noalign{\hrule}
\cr \noalign{\hrule}
\cr
Quark content &  ${\phantom ~~~}$ $R(123)$ (MeV) &
${\phantom ~~~}$ $R(132)$ (MeV) & ${\phantom ~~~}$
$R(231)$ (MeV) & \cr
\cr \noalign{\hrule}
\cr
$qqq$    & ${\phantom ~~~~~~~~}$
49 \quad (49)&
${\phantom ~~~~~~~~}$ 49 \quad (49)
& ${\phantom ~~~~~~~~}$ 49 \quad (49) \cr
$qqs$    & ${\phantom ~~~~~~~~}$
51 \quad (51)  &${\phantom ~~~~~~~~}$ 34 \quad (32)
& ${\phantom ~~~~~~~~}$ 34 \quad (32)\cr
$ssq$    & 25   & ${\phantom ~~~~~~~~}$ 35 \quad (36) &
 ${\phantom ~~~~~~~~}$ 35 \quad (36)\cr
$sss$    & 26 &   26 &  26 & \cr
$qqc$    & 54 &   12  & 12 &\cr
$qsc$    & 38 &   13 &   10 \cr
$ssc$    & 28 &   11 &   11 \cr
$qqb$    & 55 &    4 &    4 \cr
$qsb$    & 39 &    5 &    4 \cr
$ssb$    & 29 &    4 &   4 \cr
$ccq$    &  5 &  14 &  14 & \cr
$ccs$    &  6 &   12  & 12 & \cr
$qcb$    & 15 &   5  &  2 \cr
$scb$    & 13 &   5  &  3 & \cr
$bbq$    & 2  &  6  &  6 & \cr
$bbs$    & 2  &  5  &  5 & \cr

\cr \noalign{\hrule}
\cr \noalign{\hrule}
}}$$

\bye